\documentclass[aps,floatfix,showpacs,twocolumn,prl]{revtex4}
\usepackage[dvips]{graphicx}
\usepackage{amsfonts}

\newcommand{\ket}[1]{{| #1 \rangle}}

\newcommand{\tr}{\mathrm{tr}\, }
\newcommand{\ave}[1]{{\langle #1\rangle}}

\begin{document}

\title{Fourier's Law in a Quantum Spin Chain and the Onset of Quantum Chaos}
\author{Carlos Mej\'{\i}a-Monasterio$^{(a)}$}
\author{Toma\v z Prosen$^{(b)}$}
\author{Giulio Casati$^{(a,c,d)}$}
\affiliation{$^{(a)}$Center for Nonlinear and Complex Systems,
Universit\`a degli Studi dell'Insubria, via Valleggio 11, Como
22100, Italy} \affiliation{$^{(b)}$Physics Department, Faculty of
Mathematics and Physics, University of Ljubljana, Ljubljana,
Slovenia} \affiliation{$^{(c)}$Istituto Nazionale per la Fisica
della Materia, Unit\`a di Como} \affiliation{$^{(d)}$Istituto
Nazionale di Fisica Nucleare, Sezione di Milano}

\date{\today}

\begin{abstract}
We  study  heat  transport in  a  nonequilibrium  steady  state of
a  quantum interacting spin chain.  We provide clear numerical
evidence of the validity of Fourier law.   The regime  of normal
conductivity is shown to set in at the transition to quantum
chaos.
\end{abstract}


\maketitle

The derivation of Fourier's law of heat conduction from the
microscopic dynamics, without any ad hoc statistical assumption,
is one of the great challenges of  nonequilibrium   statistical
mechanics \cite{lebowitz}. Even in  the context  of classical
dynamical systems, the issue  of  energy  (heat)  transport,  in
spite  of having  a  long  history (recently reviewed in
\cite{lepri}), is  not completely  settled. It has become
clear that transport theory requires that the underlying
deterministic dynamics yield a truly diffusive process. On the
other hand, it is known that classical nonlinear systems of
interacting particles, above a critical interaction strength,
typically exhibit chaotic and diffusive behaviour which then leads
to the onset of Fourier law $J=-\kappa\nabla T$, relating the
macroscopic heat flux  to the temperature gradient $\nabla T$.
Therefore deterministic chaos appears to be an essential ingredient
required by transport theory. In this perspective, of particular
interest is the problem, almost completely unexplored, of the
derivation of Fourier law from quantum dynamics. So far,
investigations have been mainly focused on linear  response
theory\cite{kubo,lrt-review,zotos,prosen98}. The
possibility instead to derive the Fourier law from quantum
dynamics, by establishing the dependence of $J$ on $\nabla T$ in a
{\em nonequilibrium steady state}, calls directly in
question the issue of quantum chaos. In this connection, a main
feature of quantum motion is the lack of exponential dynamical
instability\cite{casati}, a property which is at the heart of 
classical dynamical chaos. This fact may render very questionable the
possibility to derive the Fourier law of heat conduction in
quantum mechanics. However, quite interestingly, it has been shown
that strong, exponential unstable, classical  chaos is not
necessary \cite{li} (actually, strictly speaking, is not even
sufficient \cite{leprifpu}) for normal transport. Thus it is
interesting to inquire if, and under what conditions, Fourier law
emerges from the laws of quantum mechanics. This is the purpose of
the present paper.

To investigate this problem one has to deal with a finite open
system connected to heat baths. Here we consider an interacting
quantum spin-$1/2$ chain which exhibits the transition  from
integrability to quantum chaos as a parameter, e.g.  the magnetic
field, is varied. The standard treatment of this problem is based
on the master equation, thus limiting investigations to relatively
small system sizes. By using this method, in an interesting
paper\cite{saito}, the decay of current correlation function  in a
model of non-integrable chain of quantum spins is computed.
 Here we take a different approach
namely we follow the evolution of the system described by a  {\em
pure} state, which is {\em stochastically} coupled to an idealized
model of heat baths. Stochastic  coupling is realized in terms of
a local measurement at the  boundary of the  system and stochastic
but unitary exchange of  energy between the  system and the  bath.
By this method we have been able to perform very  effective
numerical simulations which allow to observe  a clear
energy/temperature profile and to measure the heat current $J$.
In the nonintegrable regime where the
spectral statistics is described by random matrix theory (RMT) - the
regime  of ``quantum chaos'' - we found very accurate Fourier law
scaling $J/\Delta T \propto  1/L$, where $L$ is  the size of the
chain. In the integrable and near-integrable regimes instead, we
found that the heat transport is ballistic $J \propto L^0$.

We  consider an  Ising chain  of $L$  spins $1/2$  with coupling  constant $Q$
subject  to  a uniform  magnetic  field  $\vec{h}  = (h_x,0,h_z)$,  with  open
boundaries. The Hamiltonian reads
\begin{equation} \label{eq:H}
{\mathcal H} = -Q\sum_{n=0}^{L-2}\sigma^z_n\sigma^z_{n+1} +
\vec{h}\cdot\sum_{n=0}^{L-1}\vec{\sigma}_n \ ,
\end{equation}
where the operators  $\vec{\sigma}_n =
(\sigma^x_n,\sigma^y_n,\sigma^z_n)$ are the Pauli matrices for the
$n$-th spin, $n=0,1,\ldots L-1$.  We set the coupling constant
$Q=2$. In this system, the only trivial symmetry is a reflection
symmetry, $\vec{\sigma}_n\rightarrow\vec{\sigma}_{L-1-n}$.
Moreover  the direction of  the magnetic field affects the
qualitative behavior of the  system: If $h_z=0$, the Hamiltonian
(\ref{eq:H}) corresponds to the Ising chain  in a transversal
magnetic field. In this case the  system is integrable as
(\ref{eq:H}) can be mapped  into   a  model  of  free  fermions
through  standard Wigner-Jordan transformations.  When $h_z$  is
increased from zero, the  system is no longer integrable.  When
$h_z$ is of the same  order of $h_x$ quantum  chaos sets in
leading to a very complex  structure  of quantum  states as well
as to fluctuations in the spectrum that are statistically
described by RMT\cite{rmt}.  The system
becomes again  (nearly) integrable when $h_z \gg h_x$. Therefore,
by  choosing the direction of the external field we can explore
different  regimes of quantum dynamics.

The  onset of  quantum  chaos is  commonly  studied in  terms  of
the  nearest neighbor level spacing distribution  $P(s)$ that
gives the probability density to find two  adjacent levels at a
distance $s$.  For  an integrable system the distribution   $P(s)$
has   typically   a   Poisson   distribution   $P_{\rm
P}(s)=\exp\left(-s\right)$.  In contrast,  in  the quantum  chaos
regime (for Hamiltonians obeying  time-reversal invariance),
$P(s)$ is given  by the Gaussian Orthogonal  Ensemble of random
matrices (GOE).  In this case, $P(s)$ is well-approximated by the
Wigner surmise $P_{\rm WD}(s) =  (\pi s/2)\exp(-\pi s^2/4)$, 
exhibiting the so-called ``level repulsion''.

\begin{figure}[!t]
\begin{center}
\includegraphics[scale=0.43]{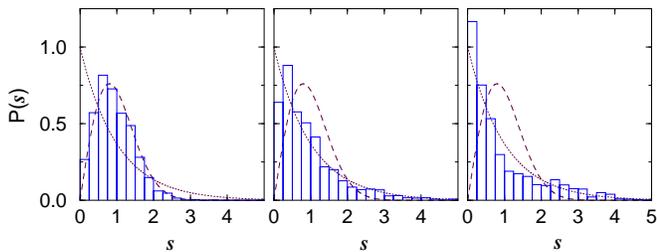}
\caption{\label{fig:1} Nearest neighbor  level spacing distribution $P(s)$ for
the  chaotic  (left), intermediate  (center)  and  integrable (right)  chains.
$P(s)$ was obtained by diagonalizing  the Hamiltonian (\ref{eq:H}) for a chain
of length  $L=12$, averaging over even  and odd parity  subspectra.  The curves
correspond to $P_{\rm WD}$ (dashed line) and to $P_{\rm P}$ (dotted line).}
\end{center}
\end{figure}

In Fig.~\ref{fig:1} we show the results of our numerical
simulations of system (\ref{eq:H}) 
 for three different values   of  the   magnetic field:
({\it   i}) {\em   chaotic   case} $\vec{h}=(3.375,0,2)$ at which
the distribution $P(s)$ agrees with GOE  and thus corresponds to
the regime of quantum chaos, ({\it ii}) {\em integrable case}
$\vec{h}=(3.375,0,0)$, at which $P(s)$ is close to the Poisson
distribution, and ({\it iii}) {\em intermediate case}
$\vec{h}=(7.875,0,2)$ at which the distribution $P(s)$ shows  a
combination of (weak) level repulsion  and an exponential  tail.
We have consistently observed signatures
for the  onset of quantum chaos also in other spectral correlations as 
well as in the structure of eigenfunctions.

Let us now turn to study  energy transport  in this model system.
To this end we need to  couple  both  ends of  the  chain of spins
to thermal reservoirs at temperature $T$.  We have devised a
simple way to simulate this coupling, namely  the state of  the
spin  in contact with  the bath is statistically determined by  a
Boltzmann  distribution with parameter $T$. Our model for the
reservoirs is  analogous to the stochastic thermal reservoirs used
in classical simulations\cite{larralde} 
and we thus call it a \emph{quantum
stochastic reservoir}. We use units in which Planck and Boltzmann
constants are set to unity $\hbar=k_{\rm B}=1$.

In the representation basis of $\sigma^z_n$  the wave function at time $t$ can
be written as
\begin{equation}
\ket{\psi(t)}=\!\!\!\!\sum_{s_0,s_1,\ldots,s_{L-1}}\!\!\!\!\!\! C_{s_0,s_1,\ldots,s_{L-1}}(t)
\ket{s_0,s_1\ldots s_{L-1}}\ ,
\end{equation}
where $s_n =0,1$ represents  the \emph{up},  \emph{down} state  of
the $n$-th spin, respectively.  
The wave function at time $t$  is obtained from
the unitary  evolution operator $\mathrm{U}(t) =
\exp(-i\mathcal{H}t)$.  The interaction with  the reservoir is not
included in the  unitary  evolution.  Instead, we  assume  that
the  spin chain  and  the reservoir interact only at discrete
times with period $\tau$ at which the states of the leftmost ($s_0$)
and the rightmost ($s_{L-1}$) spins are stochastically reset. Thus,
the evolution of the wave function from time $t$ to time $t+\tau$
can be represented as
\begin{equation}
\ket{\psi(t+\tau)} = \Xi(\beta_{\rm l},\beta_{\rm r})\mathrm{U}(\tau)\ket{\psi(t)} \ ,
\end{equation}
where $\Xi(\beta_{\rm l},\beta_{\rm r})$  represents the unitary stochastic  action of the
interaction with the left  and right reservoirs at temperatures $\beta_{\rm l}^{-1}$
and $\beta_{\rm r}^{-1}$ respectively.

The action of $\Xi(\beta_{\rm l},\beta_{\rm r})$ takes place in several steps:

\noindent  ({\it  i})  The  wave  function  is  first  rotated  by  the  angle
\mbox{$\alpha  =  \tan^{-1}(h_x/h_z)$} to  the  eigenbasis  of the  components
$\sigma_\mathrm{l} = \vec{h}\cdot\vec{\sigma}_0/h$, $\sigma_\mathrm{r}
=  \vec{h}\cdot\vec{\sigma}_{L-1}/h$ of  the edge  spins along  the  field, 
that is
$\ket{\psi}  \rightarrow e^{-i\alpha(\sigma_0^y+\sigma_{L-1}^y)/2}\ket{\psi}$.
Here, $h=|\vec{h}|$ stands for the strength of the magnetic field.

\noindent   ({\it    ii})   A    local   measurement   of    the   observables
$\sigma_\mathrm{l}$,  $\sigma_\mathrm{r}$ is performed.  Then the 
state  of the
spins  at  the  borders   collapses  to  a  state  ($s_0^*$,$s_{L-1}^*$)  with
probability
\begin{equation}
p(s_0^*,s_{L-1}^*)=\sum_{s_1,\ldots,s_{L-2}}
|C_{s_0^*,s_1,\ldots,s_{L-2},s_{L-1}^*}|^2 \ .
\end{equation}
In    other   words    this    means   that    we    put   all    coefficients
$C_{s_0,s_1,\ldots,s_{L-1}}$  with $(s_0,s_{L-1})  \neq  (s_0^*,s_{L-1}^*)$ to
zero.

\noindent  ({\it iii})  The new  state of  the edge  spins  is
stochastically chosen. After this, which simulates the thermal interaction
with  the
reservoirs, each  of the edge spins is set to {\em down}, ({\em up}) 
state with propability $\mu$,($1-\mu$).
The probability $\mu(\beta)$ depends on the
canonical temperature of each of the thermal reservoirs:
\begin{equation} \label{eq:canonical}
\mu(\beta_j)  =   \frac{e^{\beta_jh}}{e^{-\beta_jh}  +  e^{\beta_jh}}
\quad ; \quad j \in \{\mathrm{l,r}\} \ .
\end{equation}

\noindent  ({\it  iv}) Finally,  the  wave function  is  rotated  back to
the
$\sigma^z_n$              
basis, $\ket{\psi} \rightarrow
e^{i\alpha(\sigma_0^y+\sigma_{L-1}^y)/2}\ket{\psi}$.

This completes the description of  the interaction with the
quantum stochastic bath. This interaction thus
(periodically) resets the value of the local  energy
$h\sigma_{\rm l,r}$ of the  spins in contact with the
reservoirs. This information is then transmitted to the rest of
the system during its dynamical evolution and relaxation towards
equilibirum. Therefore, the
value of  $\tau$ controls the strength of the  coupling to the
bath. We have found that, in our units, $\tau=1$
provides an optimal choice.  We have nevertheless performed
simulations for other values of $\tau$ with qualitatively similar
results. 

\begin{figure}[!t]
\begin{center}
\includegraphics[scale=0.4]{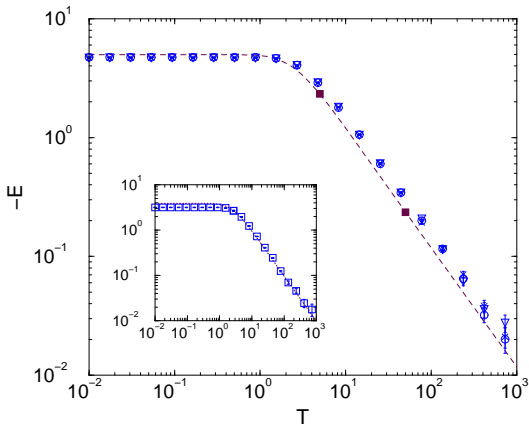}
\caption{\label{fig:2}  Mean local energy $E$ as a function of the
temperature $T$ for the chaotic  chain, for $L=6,8,10$ with crosses, triangles
and circles, respectively. 
Each point corresponds to the equilibrium
simulation in which the bath temperatures are set to the same value $T$. 
The average was obtained from the center of the profile
so that, the two sites closest  to each border were not considered. The dashed
curve corresponds  to the canonical average $E_{\rm can}(T)$
of  the local 2-body energy  operator $H_n$.
The two solid  squares indicate  the  values  of  the
temperature that were later used in  the out of equilibrium simulations.
In the inset the same data are shown for the integrable chain (squares) for
$L=8$.}
\end{center}
\end{figure}

The unitary  evolution $U(t)\ket{\psi}$ of the system has been
computed by an accurate high order split-step factorization of the
unitary evolution operator \cite{tralala}.
For each  run the initial  wavefunction $\ket{\psi(0)}$ 
of  the system is chosen
at random.   The  system  is  then evolved for some relaxation
time $\tau_{\rm rel}$ after which it is assumed to fluctuate
around a unique steady  state. Measurements are then performed  as
time averages  of  the expectation value of suitable observables.
We further average these quantities over different random
realizations of ``quantum trajectories''.

\begin{figure}[!t]
\begin{center}
\includegraphics[scale=0.4]{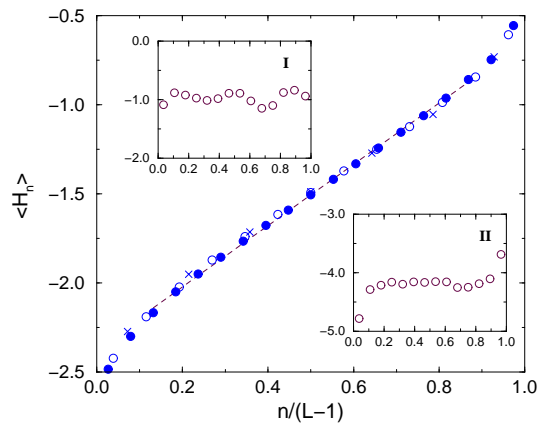}
\caption{\label{fig:3} Energy  profile $\ave{H_n}$ for an out of equilibrium 
situation for the chaotic chain.   The nominal values of the temperatures of
the reservoirs  $T_{\rm l} = 5$  and $T_{\rm r}  = 50$, are  both in the  high temperature
regime  (see the  solid  squares in  Fig.~\ref{fig:2}.  The different  symbols
correspond to chains of size $L=8$ (crosses), $L=14$ (open circles) and $L=20$
(solid circles).  The  dashed line was obtained from a linear  fit of the data
for  $L=20$  in  which  the  two   sites  closest  to  each  border  were  not
considered. In  the insets (I)  and (II) the  energy profile is shown  for the
integrable and intermediate cases respectively, for $L=15$.}
\end{center}
\end{figure}

In  order  to   compute  the  energy  profile  we   write
Hamiltonian (\ref{eq:H}) in terms of  local energy density
operators $H_n$:
\begin{equation} \label{eq:H_local}
H_{n}     =   -Q\sigma^z_n\sigma^z_{n+1}    +
\frac{\vec{h}}{2} \cdot \left(\vec{\sigma}_n  + \vec{\sigma}_{n+1}\right) \ .
\end{equation}
The local Hamiltonian $H_n$ (defined for $0 < n < L-2$), gives the
energy density   between   the   $n$-th   and   $(n+1)$-th spins.
In   terms   of eq.~(\ref{eq:H_local}) the Hamiltonian of the
system can be rewritten as
\begin{equation} \label{eq:Htot}
\mathcal{H} = \sum_{n=0}^{L-2}H_n +
\frac{h}{2}(\sigma_\mathrm{l} + \sigma_\mathrm{r}) \ .
\end{equation}

First  we have  performed  equilibrium  simulations in order to
show that time averaged expectation values of the local energy
density can be used as a consistent canonical local temperature.
To this end we set the left and  right reservoirs to the same
temperature $T$.   
In Fig.~\ref{fig:2}  we plot the mean local energy $E=\ave{H_n}$ where the 
average
is taken over time and over the $L-5$ central values of the
profile, for the chaotic and for the integrable chain (inset).
For low $T$, $E$
saturates to a constant which, together with the energy profile 
$\ave{H_n}$, is determined by the ground state.
However, for larger $T > 1$, the energy profile is
constant within numerical accuracy, and numerical simulations give
$E \sim -1/T$, all  results being  almost independent of $L$ for
$L\ge 6$. The numerical data for $E(T)$ can be well approximated with a
simple  calculation of  energy density for a two-spin chain
($L=2$) in a canonical  state at temperature $T$ (smooth curves), namely
$E_{\rm can}(T)=\tr H_0 e^{-H_0/T}/\tr e^{-H_0/T}$.
Therefore, if the temperatures of both reservoirs are in high $T$ regime,
then we can define the local temperature via the relation $T \propto -1/E$.
We stress that equilibrium numerical data shown are insensitive to the
nature of dynamics (consistent with results of Ref.\cite{jensen}), 
whether being chaotic, regular or intermediate.
\begin{figure}[!t]
\begin{center}
\includegraphics[scale=0.4]{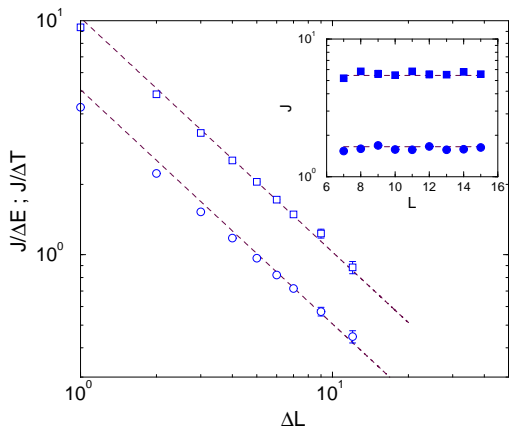}
\caption{\label{fig:4} Size  dependence of the energy  current 
in the chaotic chain with 
$T_{\rm l} = 5$ and $T_{\rm r} = 50$.  
We show $J/\Delta E$  (open circles) and $J/\Delta  T$ (open squares).
The dashed lines corresponds to  $1/\Delta L$ scaling.  
In the inset, the size dependence of the energy current is
shown for the integrable (solid circles) and the intermediate (solid squares).
}
\end{center}
\end{figure}


In Fig.~\ref{fig:3}  we show the energy
profile $\ave{H_n}$ for an out  of equilibrium simulation of the chaotic chain.
In all non-equilibrium simulations, the temperatures of the baths were set to
$T_{\rm l}=5$ and  $T_{\rm r}=50$.  
After an appropriate scaling the  profiles for different sizes collapse to the
same curve. More interesting, in the bulk of the chain the energy profile is
in very good approximation linear.  
In contrast, we  show that in
the case  of the integrable (inset  I) and intermediate (inset  II) chains, no
energy gradient  is created.  

We now define the local  current  operators  through  the  equation  of 
continuity:
$\partial_t{H_n}  = i[{\mathcal  H},H_n]  =  -  (J_{n+1} -  J_n)$,
requiring  that $J_n  =  [H_n,H_{n-1}]$.  Using  eqs.  (\ref{eq:H_local})  and
(\ref{eq:Htot}) the local current operators are explicitly given by
\begin{equation}
J_{n} = h_xQ\left(\sigma_{n-1}^z-\sigma_{n+1}^z\right)\sigma^y_{n},
\quad
1\le n\le  L-2.
\end{equation}
\begin{figure}[!t]
\begin{center}
\includegraphics[scale=0.4]{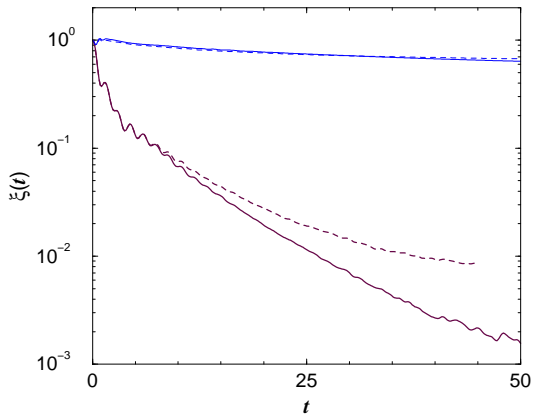}
\caption{\label{fig:5}  Infinite temperature current-current  time correlation
function $\xi(t)$, for  two chaotic chains of size $L=18$  (tick dashed line) 
and $L=24$ (thick solid line) and for two integrable chains of size $L=18$ 
(dashed line) and  $L=24$ (solid line). 
}
\end{center}
\end{figure}

In Fig.~\ref{fig:4} we plot $J/\Delta E$ as a function of the size $L$ of the
system for sizes up to $L=20$.  The mean current $J$ is calculated as an
average of $\ave{J_n}$ over time and $L-8$ central values of $n$.  The energy
difference was obtained from the energy profile as $\Delta E = \ave{H_{L-5}} -
\ave{H_3}$.  Three spins near each bath have been discarded in order to be in
the bulk regime. Since $\Delta L=L-8$ is an effective size of the truncated
system, the observed $1/\Delta L$ dependence confirms that the transport is
normal.  Moreover, also the quantity $J/\Delta T$, where $\Delta T =
-1/\ave{H_{L-5}} + 1/\ave{H_3}$, shows the correct scaling with the size $L$.
On the other hand, in integrable and intermediate chains we have observed that
the average heat current does not depend on the size $J\propto L^0$, clearly
indicating a ballistic transport.

As a consistency  check, we have also studied  the energy current-current time
correlation   $\xi(t)   =   \ave{J(t)J}$    of   the   isolated   system   for
$\beta\rightarrow  0$ which, for  high temperature, gives the  conductivity as
$\kappa=\beta  L\int_0^\infty\xi(t)\mathrm{d}t$.  In Fig.~\ref{fig:5}  we show
that for the chaotic chain  $\xi(t)$ exhibits fast, possibly exponential decay
while for the intermediate case we observe slow decay, possibly to a finite
plateau, giving rise to ballistic transport.
In the integrable case  $\xi(t)$ is constant  since $J(t)$ is a conserved
quantity.


In this paper we have shown that Fourier law of heat conduction can be
derived from the pure quantum dynamical evolution without any additional 
statistical assumption. This possibility is strictly related to the
onset of quantum chaotic behavior.

\begin{acknowledgments}
TP acknowledges financial support by the
grant P1-0044 of the Ministry of Science and Technology of Slovenia. 
\end{acknowledgments}


\begin{thebibliography}{}
\bibitem{lebowitz}F.   Bonetto,  J.L.  Lebowitz  and  L.   Rey-Bellet in  {\it
Mathematical  Physics  2000},  A.    Fokas,  A.   Grigoryan,  T.   Kibble  and
B. Zegarlinsky (Eds.), Imperial College London, 128 (2000).

\bibitem{lepri} S. Lepri, R. Livi and  A. Politi {\it Phys.Rep.} {\bf 377}, 1
  (2003).

\bibitem{li} B. Li, G. Casati, J.  Wang and T. Prosen, {\it Phys. Rev. Lett.}
{\bf 92}, 254301 (2004).

\bibitem{leprifpu}S. Lepri,  R. Livi  and A. Politi,  {\it Phys.  Rev. Lett.}
{\bf 78}, 1896 (1997).

\bibitem{kubo} R. Kubo, {\it J. Phys. Soc. Jpn.} {\bf 12}, 570 (1957).

\bibitem{lrt-review} X. Zotos and P. Prelov\v sek, {\tt cond-mat/0304630}.

\bibitem{zotos} X. Zotos, F. Naef and P. Prelov\v sek, {\it Phys. Rev. B}, {\bf
  55}, 11029 (1997)

\bibitem{prosen98} T. Prosen, {\it Phys. Rev. Lett.}, {\bf 88}, 1808 (1998)

\bibitem{casati} G.~Casati, B.~V.~Chirikov, I.~Guarneri and
D.~L.~Shepelyansky, {\em Phys. Rev. Lett.} {\bf 56}, 2437 (1986).


\bibitem{saito} K. Saito, {\it Europhys. Lett.} {\bf 61}, 34 (2003).

\bibitem{rmt}  T.   Guhr,  A.   M\"uller-Groeling, and  H.A.   Weidenm\"uller,
{\it Phys. Rep.} {\bf 299}, 189 (1998)


\bibitem{larralde}   H.  Larralde,  F.   Leyvraz,  C.   Mejia-Monasterio  {\it
J. Stat. Phys.} {\bf 113}, 197 (2003).

\bibitem{tralala} On the basis of a general idea of M. Suzuki,
Phys. Lett. A {\bf 165}, 387 (1992),
we have derived and used a very convenient complex-coefficient 
split-step formula for the short-time
evolution $U(\epsilon)=e^{(A+B)\epsilon}=e^{p_1 A\epsilon}
e^{p_2 B\epsilon} e^{p_3 A\epsilon} e^{p_4 B\epsilon} e^{p_5 A\epsilon}
+ {\cal O}(\epsilon^4)$, with $p_1=p_5^*=(1+i/\sqrt{3})/4$,
$p_2=p_4^*=2 p_1$, $p_5=1/2$.

\bibitem{jensen}R.~V.~Jensen and R.~Shankar, {\it Phys. Rev. Lett.} 
{\bf 54}, 1879 (1985).


\end{thebibliography}
\end{document}